\documentclass[a4]{elsart5p}

\usepackage{amssymb}

\usepackage{epsfig}
\usepackage{graphicx}
\usepackage{epstopdf}
\usepackage{psfrag}
\usepackage{bm}
\usepackage{color}
\usepackage{amsmath,amsfonts} 
\usepackage{tabularx}

\newcommand{\be}{\begin{eqnarray}}
\newcommand{\ee}{\end{eqnarray}}

\newcommand{\mbf}[1]{{\mathbf{#1}}}

\journal{Physics Letters B}

\begin{document}
\topmargin -1cm
\newlength{\axnucl}

\begin{frontmatter}



\title{Interplay of single-particle and collective modes in the $^{12}$C(p,2p) reaction near 100 MeV}


\author[ITPA]{A.~Deltuva},
\author[DFFCUL,CFTC]{E.~Cravo},
\author[IST,C2TN]{R.~Crespo},
\author[ITPA]{D. Jur\v{c}iukonis}

\address[ITPA]{Institute of Theoretical Physics and Astronomy, Vilnius University, 
Saul\.etekio al. 3, LT-10257  Vilnius, Lithuania}

\address[DFFCUL]{Departamento de F\'{\i}sica, Faculdade de Ci\^encias, Universidade de Lisboa, Campo Grande, 1749-016~Lisboa, Portugal}

\address[CFTC]{Centro de F\'{\i}sica Te\'{o}rica e Computacional, Faculdade de Ci\^encias, Universidade de Lisboa, Campo Grande, 1749-016~Lisboa, Portugal}

\address[IST]{Departamento de F\'{\i}sica, Instituto Superior T\'ecnico, Universidade de Lisboa, Av.~Rovisco~Pais~1, 1049-001, Lisboa, Portugal}

\address[C2TN]{Centro de Ci\^encias e Tecnologias  Nucleares, Universidade de Lisboa,
Estrada Nacional 10, 2695-066 Bobadela, Portugal}

\begin{abstract}
The $^{12}$C(p,2p)$^{11}$B  reaction at $E_p$=98.7 MeV proton beam energy is analyzed using
a rigorous three-particle scattering formalism extended to include the internal excitation of 
 the nuclear core or residual nucleus. The excitation proceeds via the core
interaction  with any of the external nucleons. We assume the $^{11}$B ground and low-lying excited states 
[$\frac32^-$ (0.0 MeV), $\frac52^-$ (4.45 MeV), $\frac72^-$ (6.74 MeV)] and 
  the excited states  [$\frac12^-$ (2.12 MeV), $\frac32^-$ (5.02 MeV)] to be members of 
$K=\frac32^-$ and  $K=\frac12^-$ rotational bands, respectively. 
The dynamical core excitation results in a significant cross section for  the reaction  
leading to the $\frac52^-$ (4.45 MeV) excited state of $^{11}$B that cannot be populated through the single-particle excitation mechanism.
 The detailed agreement between the theoretical calculations and data depends on the used optical model parametrizations
and  the kinematical configuration of the detected nucleons.

\end{abstract}




\end{frontmatter}


\section{Introduction}

The one-nucleon removal reactions  have been extensively used to study the single-particle configurations of the involved nucleus $A$ and the population of the states of its $B=(A-1)$ residue.
 At the same time, it is becoming widely accepted that single-particle, molecular/cluster and collective degrees of freedom of a nuclei coexist along the nuclear landscape. 
These  rich aspects of the nuclear structure are standardly described  by shell models, cluster and collective structure models, respectively
\cite{Caurier-05, Oertzen06, Kanada14, Bohr57}.
Concurrently, {\it ab initio} models that solve the Schr\"odinger equation for the many-body system of protons and neutrons have been developed with Hamiltonians based 
on the fundamental theory of the strong interactions \cite{Ma11} or effective interactions \cite{wiringa14}. 

Despite tremendous advances in reaction and structure, Nuclear Physics has been indulging in the artificial separation between these two branches, notwithstanding one aims to extract nuclear properties from reactions.
Some progress have been made recently to fully describe bound and scattering states or to some extent incorporate a many-body description of the nucleus into reactions 
\cite{Navratil08,Barranco17,Grinyer-12,Me19,Cr20,Rotureau20}. 
Within the later approach,  {\it ab initio} shell-model-like Variational Monte Carlo  (VMC) wave functions  (WFs) \cite{wiringa14} were recently
employed to model the single-nucleon removal  from light nuclei \cite{Grinyer-12,Me19,Cr20} under the inert-core assumption where  the knockout/breakup operator does not change the internal structure of the core.
 As a result of this crucial assumption,  the  one-nucleon spectroscopic overlap defined as a projection of the parent nucleus $A$ state onto an antisymmetrized  core + valence nucleon  $(B+N)$ form  
\cite{Cr20,brida11} becomes a key structure input for the reaction formalism. 
For a given state of the residual nucleus, the  one-nucleon spectroscopic overlap is a superposition of different nucleon angular momentum channels, $\ell j$, satisfying the appropriate triangular relations \cite{brida11}. 
The strength of the overlap or the so called spectroscopic factor (SF) for a given transition is obtained from the integral of the one-nucleon overlap function in each angular momentum channel.
The systematic study of p-shell nuclei via the single-nucleon removal channel reactions has shown that  the VMC WFs may overpredict the experimental data by almost a factor of two  for  light systems \cite{Grinyer-12}. 
This discrepancy raises a question on the  reaction model, in particular, the validity of the underlying inert-core assumption. 
Furthermore, this reaction model is unable to predict the cross sections for the residue $B$ states 
 absent in the initial nucleus $A$, i.e., those with (almost) vanishing overlap and SF.

The $^{12}$C isotopes and isotones constitute a paradigmatic light nuclei sample in which it is possible to study the coexistence of different structure aspects and its signature in reactions.  
 {\it Ab initio} VMC   WFs for $^{12}$C were recently
employed in benchmark calculations of neutrinoless double beta decay~\cite{Wang-19}  and for $^{11}$B in studies of nuclear charge radii of boron
isotopes~\cite{Maab-19}. 
Total cross sections, angular and energy  distributions, and polarization observables  for {$^{12}\mathrm{C}(p,2p)$} were measured  at GSI \cite{Panin16} and RCNP \cite{Kawase18} in inverse  and direct  kinematics, respectively, 
around 400 MeV/A energy
for the ground state $\frac32^-$ and low-lying excited states 
$\frac12^-$ (2.12 MeV) and $\frac32^-$ (5.02 MeV) of $^{11}\mathrm{B}$.
The VMC WFs  were  used to model  the proton-removal reaction $(p,2p)$  \cite{Me19,Cr20}  using standard few-body reaction frameworks where the final residue remains inert during the scattering process.
It was found that the VMC spectroscopic strength appears to be distributed among the low lying states differently than the deduced experimental values, and that the agreement between the data and predictions using these {\it ab initio} WFs diminishes prominently for transitions to excited states of $^{11}\mathrm{B}$ \cite{Me19,Cr20}.

Furthermore,  the {$^{12}\mathrm{C}(p,2p)$} reaction was also measured at a lower energy of 98.7 MeV in direct kinematics at the Indiana University 
Cyclotron Facility (IUCF), capable to separate the ground $\frac32^-$ and low-lying excited 
$\frac12^-$ (2.12 MeV), $\frac52^-$ (4.45 MeV) and $\frac32^-$ (5.02 MeV) negative-parity  states of $^{11}\mathrm{B}$  \cite{Devins-79}. 
 The outgoing protons were detected in a coplanar geometry, in two different geometries around the Quasi Free Scattering (QFS) or no-core-recoil condition.
This experiment has shown a strong population of the $\frac52^-$ (4.45 MeV) state that cannot be understood from the dominance of the single-particle knockout. 

Meanwhile, single particle and collective aspects of nuclear structure  have been incorporated in few-cluster nuclear reactions.  
The collective mode, simulated as a dynamical excitation of the nuclear core, was found  to play an important and characteristic role
in  three-cluster breakup reactions \cite{Moro-12-A, Moro-12-B, Summers-07,Diego-14,Diego-17,Deltuva-13,Deltuva-17,Deltuva-19}.

In this manuscript we aim  to reanalyze the IUCF data taking into account dynamical excitation of the  $^{11}\mathrm{B}$ core during the scattering process and get insight in the $(p,2p)$  reaction mechanisms.
In particular, whether the core excitation mechanism can be responsible for the transitions to $^{11}\mathrm{B}$ states with vanishing SF in the initial
$^{12}\mathrm{C}$.

\section{Formalism}
\label{sec2}

We use the Faddeev formalism  \cite{faddeev:60a} for three-particle scattering,
but extended to include the internal excitation of  the nuclear core, i.e., the
 residual nucleus  $(A-1)$, labeled $B$ for the brevity. The excitation proceeds via the core
interaction  with any of the external nucleons.
We work with generalized three-body
transition operators  of  Alt, Grassberger, and Sandhas (AGS)
\cite{alt:67a}. Below we shortly recall the basic equations,
whereas a more detailed description can be found in earlier works
\cite{Deltuva-19}. 

We use the usual odd-man-out notation, where, for example,
the channel $\alpha=1$ implies the particle 1 being a spectator while
particles 2 and 3 build a pair; Greek subscripts
are used for this notation. Since the nuclear core $B$
can be excited or deexcited when interacting with nucleons,
we introduce additional Latin superscript labels for the
internal state of the core, either ground (g) or excited (x).
The two-particle potentials $v_{\alpha}^{ba}$,
the two-body transition operators
\begin{equation}  \label{eq:Tg}
T_{\alpha}^{ba} =  v_{\alpha}^{ba} +\sum_{c} 
v_{\alpha}^{bc} G_0^{c} T_{\alpha}^{ca},
\end{equation}
as well as the resulting three-body transition operators
\begin{equation}  \label{eq:Uba}
U_{\beta \alpha}^{ba}  = \bar{\delta}_{\beta\alpha} \, \delta_{ba} {G^{a}_{0}}^{-1}  +
\sum_{\gamma,c} \,  \bar{\delta}_{\beta \gamma} \, T_{\gamma}^{bc}  \,
G_{0}^{c} U_{\gamma \alpha}^{ca}.
\end{equation}
couple those sets of states. 
{
The operators (\ref{eq:Tg}) and (\ref{eq:Uba}) include 
simultaneously both core excitation and single-particle-like excitations,
making those contributions mutually consistent.
}
Here $\bar{\delta}_{\beta\alpha} = 1 - \delta_{\beta\alpha}$,
$E$ is the available system energy in the c.m. frame,   and
 $G_0^{a} = (E+i0-\delta_{ax}\Delta m_B - H_0)^{-1}$ is
the free resolvent that
beside the internal-motion kinetic energy operator $H_0$ contains
also the contribution of the excitation energy $\Delta m_B$.
In this formalism the two-nucleon potential $v_\alpha^{ba}$ and the respective
transition matrix $T_\alpha^{ba}$
have only diagonal components $b=a$.
The breakup operator
\begin{equation}  \label{eq:U0a}
U_{0 \alpha}^{ba}  =  \delta_{ba} {G^{a}_{0}}^{-1}  +
\sum_{\gamma,c} \,    \, T_{\gamma}^{bc}  \,
G_{0}^{c} U_{\gamma \alpha}^{ca} \, ,
\end{equation}
corresponds to the case  $\beta=0$  in Eqs.~(\ref{eq:Uba})
and, once the coupled system for $\beta=1,2,3$ is solved, 
does not require a new 
solution of Eqs.~(\ref{eq:Uba}) but is given by the  quadrature (\ref{eq:U0a}) involving
 $U_{\gamma \alpha}^{ca}$ with $\gamma=1,2,3$.

{
The physical amplitudes for the breakup process
are obtained as the on-shell matrix elements
of $U_{0 \alpha}^{ba}$ taken between initial  and final 
channel states. We label by $\alpha=1$ the initial two-cluster state, where
the proton with the relative  momentum $\mbf{q}_1$ impinges
on the  nucleus $(B+N)$, i.e.,
 $|\Phi_1 (\mbf{q}_1) \rangle = 
(|\Phi_1^g \rangle + |\Phi_1^x \rangle) | \mbf{q}_1 \rangle $.
The spectator part $| \mbf{q}_1 \rangle $ is a free wave while
the pair part $|\Phi_1^g \rangle + |\Phi_1^x \rangle$
is a solution of the Schr\"odinger equation with a real $(B+N)$ potential that
couples ground- and excited-state core components. In the core-valence bound-state partial-wave ($0^+$ in the case of $^{12}\mathrm{C}$)
the same potential has to be used also for the transition operator in Eq.~(\ref{eq:Tg}) where it generates the bound-state pole.
This special potential (see Appendix A) is taken to have Woods-Saxon form
with central and spin-orbit terms plus Coulomb, and its parameters
are adjusted such that the resulting two-body wave-function components in the $r$-space
$\Phi_1^a (r)$  reproduce (up to a factor)   the corresponding VMC spectroscopic overlaps $R^a(r)$ defined in Ref.~\cite{brida11},
i.e., $\Phi_1^a (r) = \mathcal{N}^{-1/2} R^a(r)$. The procedure is similar to the one proposed in Ref.~\cite{brida11}
but with an important extension including the $g$-$x$ channel coupling.
The normalization factor $\mathcal{N}$ is introduced to ensure that
$|\Phi_1^g \rangle + |\Phi_1^x \rangle$ is normalized to unity as required by the AGS equations (\ref{eq:Uba}).
Since the norm of the spectroscopic overlap $R^a(r)$ is by definition the respective spectroscopic factor SF($a$),
we have $\mathcal{N} = \mathrm{SF}(g) + \mathrm{SF}(x)$. 
Using the same $\mathcal{N}$  for both $g$ and $x$ components ensures that their relative weight is preserved
 as in  VMC overlaps.
Thus, the spatial and momentum distributions of the initial channel wave function
$|\Phi_1^g \rangle + |\Phi_1^x \rangle$ up to a normalization factor
closely resemble those of the  VMC overlaps.
}

We assume that in the final three-cluster channel 
$|\Phi_0^b (\mbf{p}'_\eta,\mbf{q}'_\eta) \rangle$  one
can distinguish between the ground ($b=g$) and excited ($b=x$)
states of the core.
Instead of single-particle momenta $\mbf{k}_\eta$ we use
Jacobi momenta, where  $\mbf{p}'_\eta$  ($\mbf{q}'_\eta$) labels
the Jacobi momentum within the pair (spectator relative to the pair).
The breakup channel states
 can be expressed in any of the three Jacobi sets $\eta$.
Therefore, the amplitude for the three-cluster breakup reaction with the
core nucleus in the final state $b$ has to be calculated as
\begin{equation} 
\mathcal{T}^b_1(\mbf{p}'_\eta,\mbf{q}'_\eta;\mbf{q}_1) =
\sum_{a} \langle \Phi_0^b (\mbf{p}'_\eta,\mbf{q}'_\eta)|
U_{0 1}^{ba}  |\Phi_1^a (\mbf{q}_1) \rangle.
\end{equation}
If the two nucleons are identical, the amplitude has to be properly
antisymmetrized as discussed in Ref.~\cite{deltuva:13b}.

We consider a kinematically complete  three-particle breakup experiment
where two  particles, say, $\alpha$ and $\beta$, are detected at solid angles
$\Omega_\alpha$ and $\Omega_\beta$, respectively. Measuring energy of one
particle, $E_\alpha$, in principle determines the final-state kinematics
completely, since the remaining variables are constrained by
the energy and momentum conservation (in the kinematical region of interest
in the present work the relation between kinematic variables is unique,
though in general it may have two solutions).
The corresponding fivefold  differential cross section  is
\begin{gather}  \label{eq:d5s}
\frac{d^5\sigma}{dE_\alpha d\Omega_\alpha d\Omega_\beta}
= (2\pi)^4 \, \frac{M_1}{q_1} \,
|\mathcal{T}^b_1(\mbf{p}'_\eta,\mbf{q}'_\eta;\mbf{q}_1)|^2 \,
\mathrm{fps} 
\end{gather}
with $M_1=m_1(m_2+m_3)/(m_1+m_2+m_3)$, $m_\alpha$ being the mass
of the particle $\alpha$, and the phase-space factor
\begin{gather}  \label{eq:fps}
  \mathrm{fps} = \frac{m_\alpha m_\beta m_\gamma k_\alpha k_\beta^3 }
         {|  m_\gamma k_\beta^2
     - m_\beta \mbf{{k}}_\beta \cdot (\mbf{Q} -\mbf{k}_\alpha -\mbf{{k}}_\beta) |},
\end{gather}
$\mbf{Q}$ being the total three-particle momentum.

We  solve the scattering equations (\ref{eq:Uba}) and calculate the
breakup amplitudes
(\ref{eq:U0a})   in the momentum-space  framework
following Ref.~\cite{Deltuva-19}, with a slight difference related
to the inclusion of the Coulomb force. In the system of two protons and nuclear
core the Coulomb force acts in all three pairs of particles, leading to unknown
renormalization factor in the screening and renormalization procedure.
We therefore include screened Coulomb potentials with respective strengths
for all three pairs, but do not perform the renormalization of breakup
amplitudes. Instead, we check that the cross sections to a good accuracy become
independent of the screening radius, implying that the renormalization factor
should be just a phase factor, and the screened Coulomb potential simulates well
the actual Coulomb force acting in the breakup process.
In the considered reaction the convergence is achieved with the
Coulomb screening radius around 10 fm, but even neglecting the Coulomb force
does not lead to significant changes, implying that the Coulomb force
is quite irrelevant in the present case.

\begin{figure}[!]
\begin{center}
\includegraphics[width=0.50\textwidth]{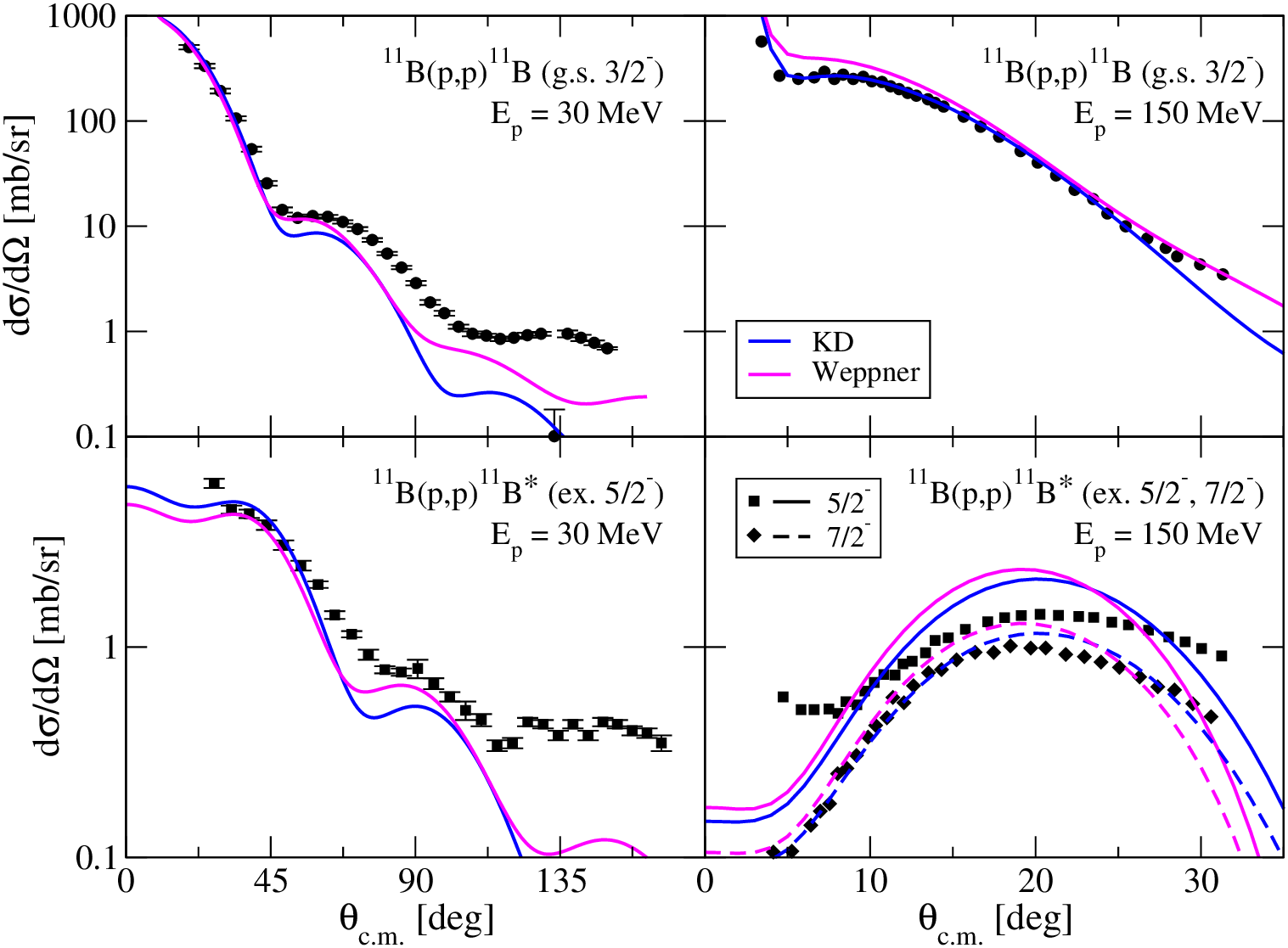}
\caption{\label{fig:inelastic11B}
Differential cross section for the  elastic and inelastic scattering $^{11}{\rm B}(p,p)^{11}{\rm B}^*$  
at $E_{p}=30.3$ MeV and 150 MeV,
leading to the member states of the $K=\frac32^-$ rotational band.
Predictions are based on 
two sets of  optical potential parametrizations, KD and Weppner.
The data is taken from   Refs.~ \cite{Karban:69} and  \cite{Hannen}.}
\end{center}
\end{figure}
%

The nucleon-residual nucleus potentials with the core excitation are constructed in a standard way using the rotational model
\cite{bohr-motelson,tamura:cex,thompson:88}.
One starts with a single-channel optical potential whose radial dependence
is usually parametrized in terms of the Woods-Saxon function. In the present
study, and to investigate the uncertainties of the calculated cross sections associated 
with the choice of the optical potential  (OP) parametrization,
we take two global OPs.  They were developed by Weppner {\it et al.} \cite{weppner}
and Koning and Delaroche (KD) \cite{Koning-03}, and fitted to
 $A \ge 12$ nuclei and $A \ge 24$ nuclei, respectively. Despite this restriction, the KD potential
has been used for systematic studies along the nuclear landscape also for lighter nuclei \cite{Cravo-16},
and reproduces the experimental nucleon-nucleus data with a reasonable quality as will be shown later.
For any of these OPs we assume a permanent quadrupole deformation of the nucleus.
This induces a coupling to the internal nuclear degrees of freedom $\hat{\xi}$ of the residual nucleus
via  the  Woods-Saxon radius $R_j = R_{j0}[1+\beta_2 Y_{20}(\hat{\xi})]$,
$\beta_2$ being the quadrupole deformation parameter,
and $\delta_2 = \beta_2 R_{j0}$ called the deformation length.
In the case here considered  it was assumed  that the ground state of spin/parity
$\frac32^-$ and excited states $\frac52^-$ (4.45 MeV) and 
$\frac72^-$ (6.74 MeV) of the  $^{11}\mathrm{B}$ residual  nucleus
are members of the $K=\frac32^-$ rotational band,
while the excited states  $\frac12^-$ (2.12 MeV) and 
$\frac32^-$ (5.02 MeV) are members of the $K=\frac12^-$ rotational band  \cite{Pham:76, Karban:69}.
{
We use axially symmetric rotational model and do not include coupling between states belonging to different
rotational bands. When looking at the experimental data for the proton + $^{11}\mathrm{B}$ inelastic scattering
\cite{Karban:69,Hannen} one notices that  $K=\frac32^-$ to  $K=\frac12^-$ 
 cross sections are 3 to 5 times lower than those within the $K=\frac32^-$ band,
which justifies our approximation.
}

The experimental data for the proton + $^{11}\mathrm{B}$ scattering,
needed to fix the potential parameters, is available in the $K=\frac32^-$ case
only, and in different energy regime of $E_{\rm p}$=30.3 MeV  \cite{Pham:76}, suggesting
 $\beta_2 \approx 0.52$, or  $\delta_2 \approx 1.5$ fm. We find this value consistent also
with the experimental data at the higher energy of  $E_{\rm p}$=150 MeV \cite{Hannen}.
Due to the lack of experimental information we use the same parameters
also for the reactions coupling the states of the 
$K=\frac12^-$ rotational band.

In summary, the used nucleon-residue potentials are based on the
Weppner's and KD  parametrizations with $\delta_2 \approx 1.5$ fm rotational quadrupole deformation length.
An exception is the core-valence interaction in the $0^+$ partial wave
where real potentials (different for each band) are used to simulate VMC spectroscopic overlaps.
As for the two-nucleon potential we verified that results
are insensitive to its choice provided it is a realistic high-precision potential such as AV18 \cite{wiringa:95a}
or CD Bonn \cite{machleidt:01a}.

\begin{figure*}[!]
\begin{center}
\includegraphics[width=0.46\textwidth]{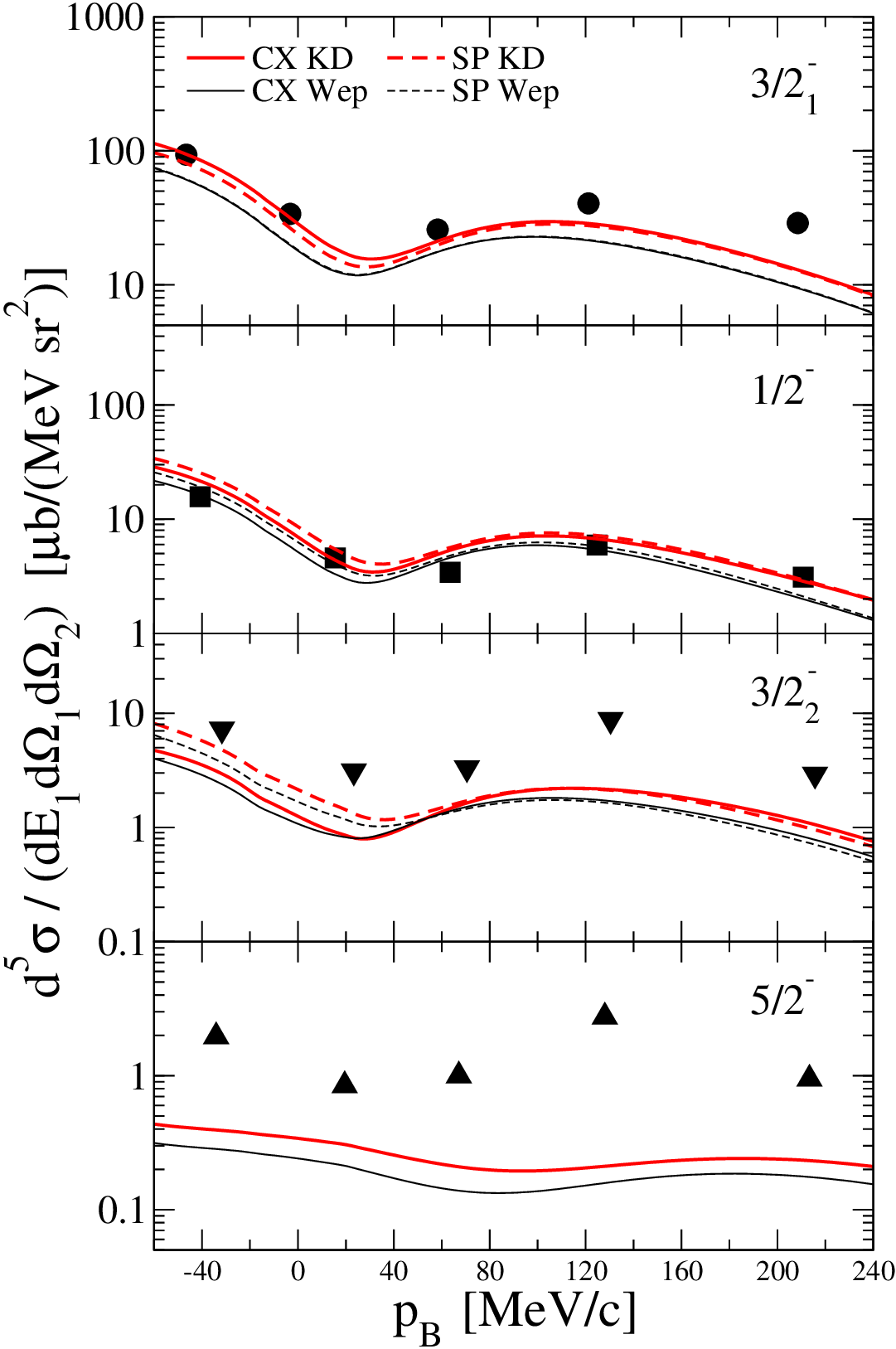}
\includegraphics[width=0.45\textwidth]{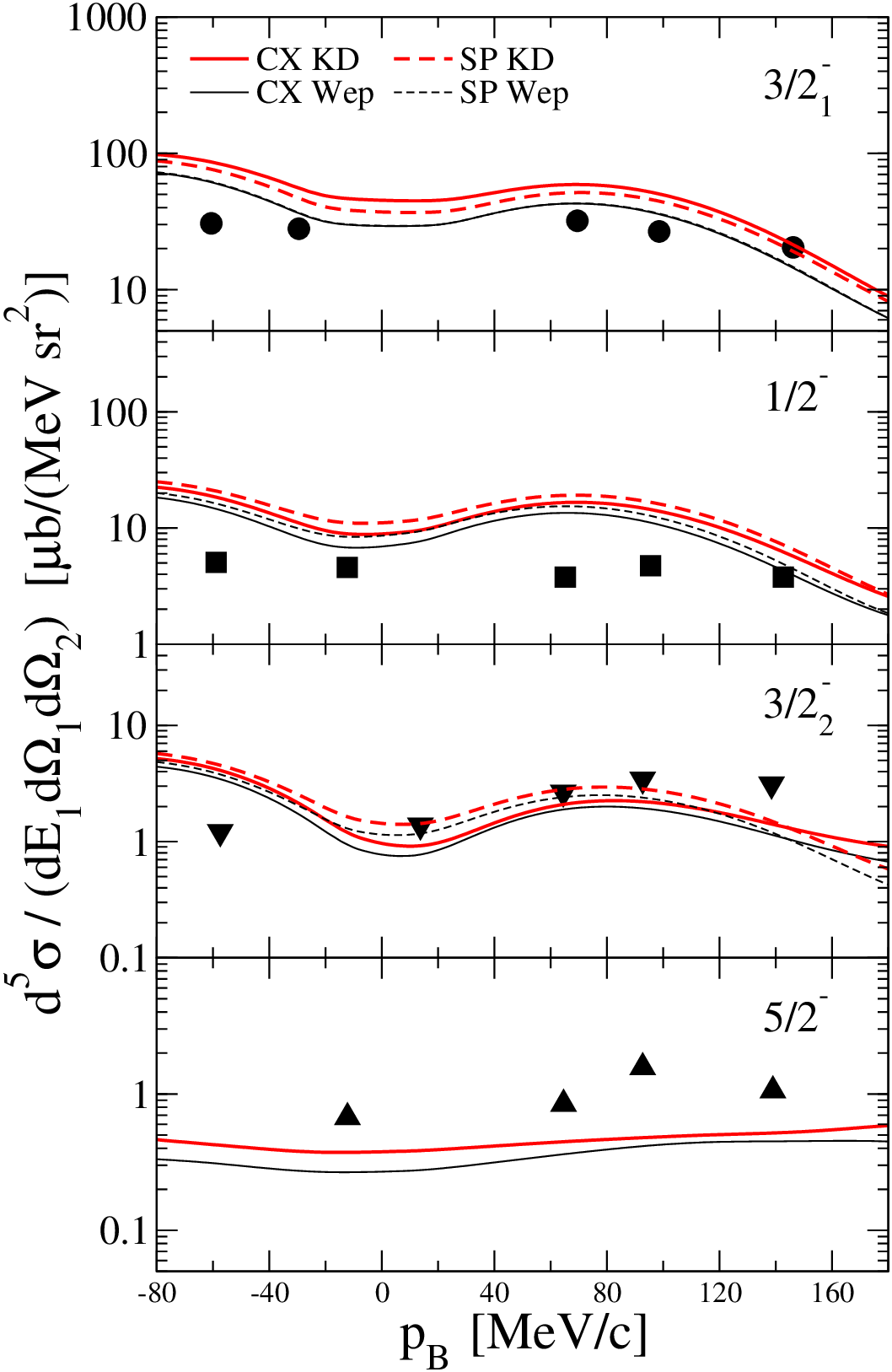}
\caption{\label{fig:Devins_CX}
Differential cross section   for the  $^{12}{\rm C}(p,2p)^{11}{\rm B}$ reaction at  $E_{p}=98.7$ MeV as function of the residual nucleus momentum $p_B$ 
in the symmetric (left) and asymmetric (right) kinematical settings of Ref.~ \cite{Devins-79}.
Predictions are based on the KD and Weppner (Wep) optical potential parametrizations,
with the dynamical core excitation either included (CX, solid curves) or neglected (SP, dashed curves). The spectroscopic VMC information is taken into account
as described in the text. The subscripts 1 and 2 distinguishe the  ground and excited $3/2^-$ states, respectively.
The data is taken from Ref.~ \cite{Devins-79}.}
\end{center}
\end{figure*}


\section{Results}

Our main goal is to study the
  {$^{12}\mathrm{C}(p,2p)$} reaction at the incident proton energy $E_p = 98.7$ MeV,  measured in direct kinematics at the Indiana University Cyclotron Facility (IUCF) \cite{Devins-79}. 
However,
we begin by testing the adequacy of the chosen nucleon-nucleus dynamical excitation model with rotational quadrupole deformation.
 Due to the lack of experimental information around $E_{p}$=100 MeV
we display in Fig.~\ref{fig:inelastic11B} the angular distributions of the differential cross section for elastic and inelatic scattering $^{11}{\rm B}(p,p)^{11}{\rm B}^*$ leading to the 
states of the $K=\frac32^-$ rotational band, and compare them with the experimental data
 at   $E_p$=30.3 MeV  and  $E_p$=150 MeV  taken from   Refs.~\cite{Karban:69} and  \cite{Hannen}, respectively. 
While at the lower energy the predictions using the Weppner's OP are somewhat closer to the data, at the higher energy the KD OP reproduces the data slightly better, 
while the Weppner OP tends to overestimate the cross sections.
Nevertheless,
the calculated proton-$^{11}$B elastic and inelastic cross sections  using  both  global OP parametrizations follow fairly well  the trend of the data
with the exception of the $\frac52^-$ state in the
forward-angle region where other multipolar transitions are expected to contribute  \cite{Hannen}.
Moreover, the data for the excitation of the  $\frac72^-$ state at 150 MeV are   well reproduced.

Next, we solve the three-body Faddeev/AGS equations including the core excitation (here labelled as CX) and
calculate the fivefold differential cross section for the $^{12}$C($p,2p)^{11}$B reaction.
In order to estimate the CX effect we performed also the corresponding calculations  without the core excitation, 
i.e., using the standard single-particle (SP) dynamic model \cite{Me19}.

The two particles detected in the Indiana experimental setup \cite{Devins-79} are the two protons, to be labeled $p$ and $N$ in the following,
though one has to keep in mind that they are indistinguishable and the scattering amplitudes are correspondingly antisymmetrized.
The emitted protons are measured in  a coplanar geometry, 
with the azimuthal angle between them being  $180^\circ$. The plane geometry reduces the number of
independent kinematical variables to three, chosen as the energy of one proton and polar angles of both protons, $\{E_p,\theta_p,\theta_N)$.
The outgoing protons were detected  in two different kinematic geometries around the Quasi Free Scattering (QFS):
A symmetric  one,   labeled KS in \cite{Devins-79}, is characterized by $E_p = 41.35 \pm 1.25$~MeV  and $\theta_p = \theta_N$
taking five values from $30^{\circ}$ to $65^{\circ}$.
 An asymmetric one, labeled KA in \cite{Devins-79}, is characterized by $E_p = 59.5 \pm 1.8$~MeV, $\theta_p = 25^{\circ}$, and $\theta_N$ taking
five values from $30^{\circ}$ to $90^{\circ}$.

{
The bound-state wave function in the three-body AGS calculations is  normalized to unity, but  multiplied by the 
normalization factor $\mathcal{N}^{1/2}$ its components reproduce the respective microscopic VMC overlaps.
 Therefore the  spectroscopic VMC information is taken into account 
multiplying  the AGS-CX  cross section results  by the norm factor  $\mathcal{N} = \mathrm{SF}(g) + \mathrm{SF}(x)$ which is the sum of 
SFs for the states coupled in the AGS calculations.
The AGS-SP calculations involve only a single state, in that case one has to multiply the cross section by the SF of the considered state,
which is a standard way to include the spectroscopic information also into the distorted-wave calculations  \cite{Me19}.
The VMC SFs for the $\frac32^-$ (0.0 MeV),
$\frac12^-$ (2.12 MeV), $\frac52^-$ (4.45 MeV), and $\frac32^-$ (5.02 MeV)
 states of $^{11}\mathrm{B}$ are 2.363, 0.819, 0.001, and 0.206, respectively \cite{Bob-overlaps}.
}

The fivefold differential cross sections, scaled by the SF's in an appropriate way, are shown in Fig~\ref{fig:Devins_CX} as functions of the core momentum $p_B$.
Results using both Weppner and KD OP parametrizations, with and without CX, for both KS (left) and KA (right) geometries are compared with the experimental data
\cite{Devins-79}.

\begin{figure}[!]
\begin{center}
\includegraphics[width=0.46\textwidth]{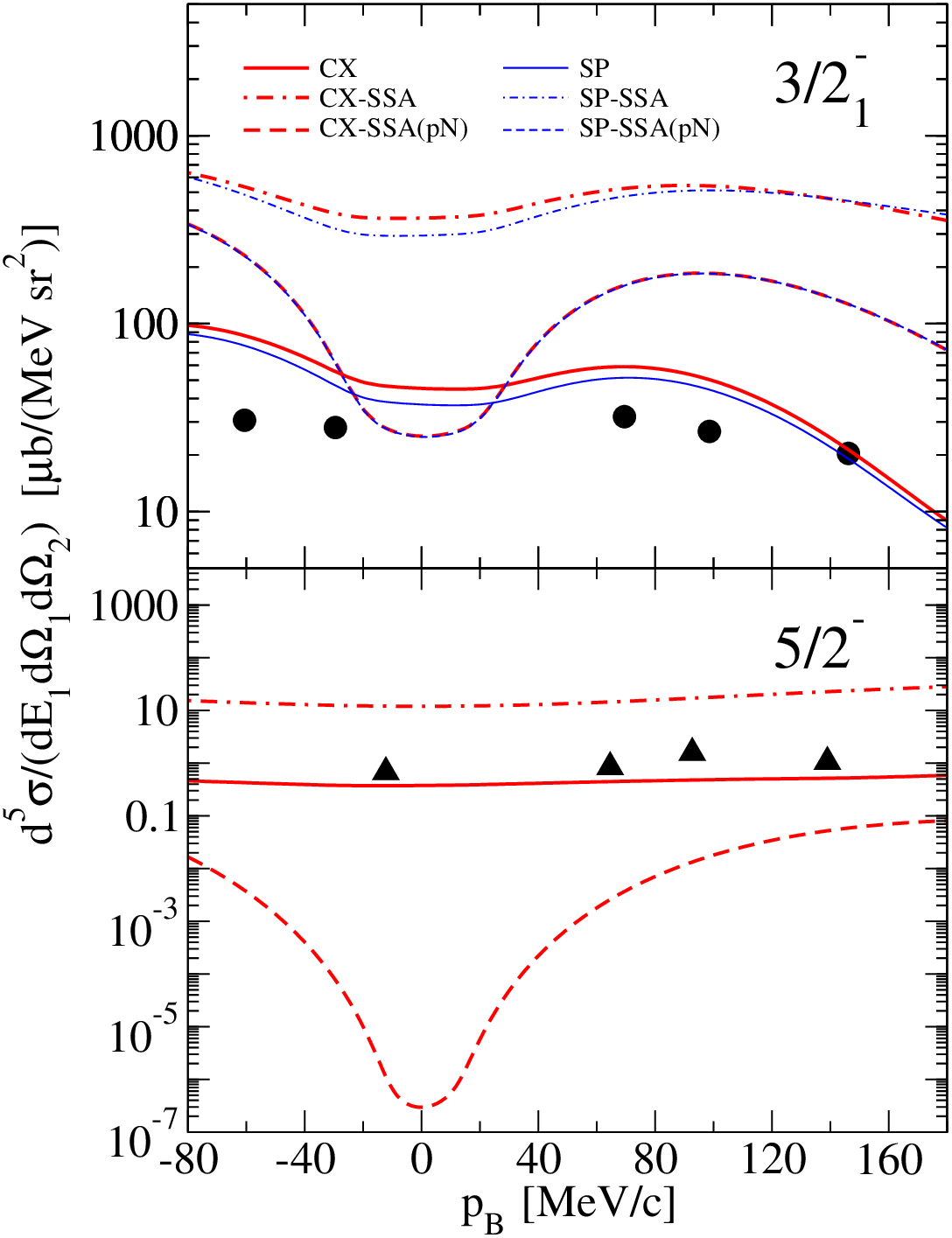}
\caption{\label{fig:ssa}
Differential cross section   for the  $^{12}{\rm C}(p,2p)^{11}{\rm B}$ reaction at  $E_{p}=98.7$ MeV as function of the residual nucleus momentum $p_B$ 
in the  asymmetric kinematical setting of Ref.~ \cite{Devins-79}.
Predictions are based on the KD  optical potential parametrization
with the dynamical core excitation  (CX, thick curves). For the $^{11}{\rm B}$ ground state results neglecting CX are also shown (SP, thin curves). 
In both cases full results are given by solid curves while SSA results with proton-core and proton-valence (only proton-valence) terms are given
by dashed-dotted (dashed) curves.
The spectroscopic VMC information is taken into account as described in the text. 
The data is taken from Ref.~ \cite{Devins-79}.}
\end{center}
\end{figure}

The general trend is that the CX effect increases the differential cros section in most parts of the considered kinematic region, CX predictions being higher than those of SP.
In the case of the $\frac52^{-}$ state the SP results are almost vanishing and are not shown.
Another trend is that KD OP parametrization leads to higher differential cros section than Weppner OP, though in the two-body case the situation is reversed
as shown in Fig.~\ref{fig:inelastic11B}. This sensitivity is slightly enhanced in the CX case.
All these features possibly indicate that the CX is a complicated phenomena resulting from interplay and partial cancellations of various terms in the dynamic equations,
or, equivalently, in the multiple scattering series. In fact, if the multiple scattering series are replaced by the first-order terms only, 
the so-called single-scattering approximations (SSA), the resulting cross section is heavily enhanced.
{ 
We illustrate this finding in Fig.~\ref{fig:ssa} by comparing full results and SSA
for transitions to the ground and $\frac52^{-}$ state in the asymmetric kinematics. Despite that SSA cross sections are much higher,
 the CX effect for g.s. is qualitatively similar in both full and SSA cases. 
For curiosity we show also the SSA(pN) including only the proton-valence term.
It excludes the proton-core interaction and thereby also the  dynamic CX,  leading to a vanishing CX effect.
In the case of the $\frac52^{-}$ state the SSA(pN) cross section is very small as a  consequence of  very small SF. 
Thus, it is indeed the dynamic CX that is responsible for an appreciable cross section for the final $\frac52^{-}$ state}.

With respect to reproducing the experimental data \cite{Devins-79} the situation is quite contradictory. The differential cros section for the 
 $^{11}\mathrm{B}$ in its ground state $\frac32^-$ is well described by the CX KD calculation in the KS kinematics and by SP and CX Weppner calculation in the KA kinematics.
For the other member of the $K=\frac32^-$ rotational band, the $\frac52^{-}$ state, the  differential cros section is significantly underestimated in the 
KS kinematics but only slightly underestimated in the  KA kinematics by the CX KD calculation.
For the $K=\frac12^-$ rotational band the transition to the excited state $\frac12^-$ is quite well reproduced by all calculations in the 
KS kinematics but overpredicted in the KA kinematics. On the contrary, the transition to the excited state $\frac32^-$(5.02 MeV) is underestimated in the
KS kinematics but, except for one point,  described well by the  KD calculations (both with and without CX)  in the  KA kinematics.

Thus, while in average the CX KD calculations appear to be more successful than the others, no single calculation provides a reasonable reproduction of all 
the experimental data \cite{Devins-79}. The quality of the description depends on the kinematics,  in the KS being better for 
the $\frac32^-$ (0.0 MeV)  and $\frac12^-$ (2.12 MeV) states,  while in KA being better for the $\frac52^-$ (4.45 MeV) and $\frac32^-$ (5.02 MeV)
 states. The reason remains unexplained.

Additionally, our models including dynamical core excitation  predict differential
 cross section  for the transition to the excited state  $\frac72^-$  to be smaller than the one for the $\frac52^-$ but still of the same order of magnitude. 
We do not show it here since the experimental data is not available for a pure state, only as a mixture with the $\frac12^+$ state.
In the same way as $\frac52^-$, the transition to $\frac72^-$ cannot be described by the SP model due to the associated negligible SF.

\section{Summary and conclusions}

We have reinterpreted the experimental data for the  $^{12}$C(p,2p)$^{11}$B reaction at $E_p=98.7$ MeV,
in which the emitted protons  are measured in  a coplanar geometry, with the azimuthal angle between them being  $180^\circ$. 
Two kinematical settings have been considered.

We used an extended  three-particle  reaction formalism that includes the internal excitation of the nuclear core. The excitation proceeds via the core
interaction with any of the external nucleons. We assume the $^{11}$B ground and low-lying excited states [$\frac32^-$, $\frac52^-$ (4.45 MeV), $\frac72^-$ (6.74 MeV)] and 
the excited states  [$\frac12^-$ (2.12 MeV), $\frac32^-$ (5.02 MeV)] to be the members of the $K=\frac32^-$ and  $K=\frac12^-$ rotational bands, respectively. 
  
A detailed agreement between the theoretical calculations and data is somehow contradictory and
depends on the used optical potential parametrization and the final-state kinematical situation. This 
possibly indicates that the core excitation is a complicated phenomena resulting from interplay and partial cancellations of various terms in the dynamic equations but also
calls for a new data. Most importantly,
the dynamical excitation of the core included in the reaction model predicts insufficient but nevertheless quite significant cross sections 
for transitions to the  $\frac52^-$  and $\frac72^-$ excited states that cannot be populated via the single-particle excitation mechanism.

Thus,  we have shown the ability to predict at least qualitatively the cross sections for states with residual nucleus components that are negligible in the initial nucleus.
This will surely contribute also to analysis of upcoming data from $^{12}$C(p,2p)$^{11}$B measurements detecting
$\frac52^-$ (4.45 MeV) and $\frac72^-$ (6.74 MeV) states of $^{11}$B, 
currently under study at GSI and other laboratories.



\vspace{1mm}

We thank R. B. Wiringa for providing the overlap functions.
A.D. and D.J. are supported by Lietuvos Mokslo Taryba
(Research Council of Lithuania) under Contract No.~S-MIP-22-72.
Part of the computations were performed using the infrastructure of
the Lithuanian Particle Physics Consortium.

\begin{appendix}
  \section{Nuclear binding potential}
  
  We start with an undeformed valence-core potential for the $0^+$ state
  \begin{gather}  \label{eq:Vb}
\begin{split}
v_{1}(r) = & {} -V_c \, f(r,R_c,a_c) - \mbf{s}_N\cdot \mbf{s}_B \, V_{ss}\,f(r,R_c,a_c)  \\
& + \mbf{s}_N\cdot \mbf{L} \,\,
V_{ls} \, \frac{4}{r} \frac{d}{dr}f(r,R_{ls},a_{ls}),
\end{split}
\end{gather}
  where $f(r,R,a) = [1+\exp((r-R)/a)]^{-1}$ is Woods-Saxon form factor,
  $\mbf{s}_N$ and $\mbf{s}_B$ are spins of the nucleon and nucleus, respectively, and
 $ \mbf{L}$ is the orbital angular momentum.
  For the  $K=\frac12^-$ rotational band we found that appending the central term by
  a phenomenological spin-spin contribution suggested in Ref.~\cite{amos:03a} 
  improves the fit significantly.
  To include excitation of the core, the central term is deformed in a standard way
  using $\delta_2 = 1.5$ fm.
  The parameters determined from the fit and the resulting weights
  $P_a$ of the wave-function components are collected in  Table \ref{tab:v}.

  \begin{table}[t]
    \caption{\label{tab:v}
      Binding potential parameters for $K = \frac32^-$ and $\frac32^-$ rotational bands
      with quadrupole deformation length $\delta_2 = 1.5$ fm.
      The radii $R_i$ and diffuseness $a_i$ are in units of fm,
      central strength $V_c$ is in  units of MeV and spin-orbit strength $V_{ls}$ is in
      units of MeV fm$^2$. The resulting weights of the ground- and excited state
      wave-function components are listed as well.}
\begin{tabular}{|*{10}{c|}}
  \hline
  $K$ & $R_c$ & $a_c$ &   $V_c$ &  $R_{ls}$ & $a_{ls}$ &  $V_{ls}$
  & $V_{ss}/V_c$ &  $P_g$ & $P_x$ \hfill \\ \hline
 $\frac32^-$ &  3.011 &  0.692 &  63.068 & 3.169 & 0.253 & 7.555 & 0.0 &   0.997 & \hfill 0.003 \\ 
 $\frac12^-$ & 2.636 & 0.784 &  76.274 & 2.005 & 0.608 & 8.622 & 0.248 & 0.795 & 0.205 \\ 
\hline
\end{tabular}
\end{table}

\end{appendix}




\begin{thebibliography}{10}

\bibitem{Caurier-05} E. Caurier, G. Mart{\i}nez-Pinedo, F. Nowacki, A. Poves, A.P. Zuker, Rev. Mod. Phys. 77, 427 (2005).
\bibitem{Oertzen06} W. Von Oertzen, M. Freer, Y. Kanada-Enyo, Phys. Rep. {432}, 43 (2006).
\bibitem{Kanada14} Y. Kanada-Enyo, T. Suhara, F. Kobayashi, 
Eur. Phys. J. 66, 01008 (2014).
\bibitem{Bohr57} A. Bohr and B.R. Mottelson, Collective and Invididual aspects of nuclear structure, (1957).

\bibitem{Ma11}  R. Machleidt, D.R. Entem,  Phys. Rep. 503, 1 (2011).
\bibitem{wiringa14}R. B. Wiringa, \textit{et al.},
Phys. Rev. C 89, 024305 (2014).

\bibitem{Navratil08}S. Quaglioni and P. Navratil, Phys. Rev. Lett. 101, 092501 (2008).
\bibitem{Barranco17} F. Barranco  \textit{et al.}, Phys. Rev. Lett. 119, 082501 (2017).
\bibitem{Grinyer-12}G. F. Grinyer \textit{et al.}, Phys. Rev. C 86, 024315 (2012).
\bibitem{Me19}   
 A. Mecca, E. Cravo, A. Deltuva, R. Crespo, A.A. Cowley, A. Arriaga, R.B. Wiringa, T. Noro, Phys. Lett. B 798, 134989 (2019).
\bibitem{Cr20}
R. Crespo, A. Arriaga, R. Wiringa, E. Cravo, A. Deltuva, A.Mecca, Phys. Lett. B 803, 135355 (2020). 
\bibitem{Rotureau20} J. Rotureau,G. Potel, W. Li and F.M. Nunes, J. Phys. G: Nucl. Part. Phys. 47, 065103 (2020).
\bibitem{brida11}I. Brida, Steven C. Pieper and R. B. Wiringa, Phys. Rev. C 84, 024319 (2011).

\bibitem{Wang-19}X.B. Wang \textit{et al.}, Phys. Lett. B 798,134974 (2019).
\bibitem{Maab-19}B. Maab, \textit{et al.}, Phys. Rev. Lett. 122, 182501 (2019).
\bibitem{Panin16}V. Panin, \textit{et al.}, Phys. Lett. B 753, 204 (2016).
\bibitem{Kawase18} S. Kawase \textit{et al.}, Prog. Theor. Exp. Phys. 2018, 021D01 (2018).
\bibitem{Devins-79} D.W. Devins, D.L. Friesel, W.P. Jones,  A.C. Attard, I.D. Svalbe, v.C. Officer, R.S. Henderson, B.M. Spicer and G.G. Shute, 
Aust. J. Phys., 32, 323 (1979).

\bibitem{Moro-12-A} A. M. Moro and R. Crespo, Phys. Rev. C 85, 054613 (2012).
\bibitem{Moro-12-B} A. Moro and J. A. Lay, Phys. Rev. Lett. 109, 232502 (2012).
\bibitem{Summers-07} N. C. Summers and F. M. Nunes, Phys. Rev. C 76, 014611 (2007).
\bibitem{Diego-14} R. de Diego, J. M. Arias, J. A. Lay, and A. M. Moro, Phys. Rev. C 89, 064609 (2014).
\bibitem{Diego-17}R. de Diego, R. Crespo, and A. M. Moro, Phys. Rev. C95, 044611 (2017).
\bibitem{Deltuva-13} A. Deltuva, Phys. Rev. C 88, 011601(R) (2013).
\bibitem{Deltuva-17} A. Deltuva, D. Jur\v{c}iukonis, and E. Norvai\v{s}as, Phys. Lett. B 769, 202 (2017).
\bibitem{Deltuva-19} A. Deltuva, Phys. Rev. C 99, 024613 (2019).

\bibitem{faddeev:60a}
L.~D. Faddeev, Zh.~Eksp.~Teor.~Fiz. { 39},  1459  (1960), [Sov.~Phys. JETP { 12}, 1014 (1961)].

\bibitem{alt:67a} E.~O. Alt, P. Grassberger, and W. Sandhas, Nucl.~Phys. { B2},  167  (1967).

\bibitem{deltuva:13b} A. Deltuva, Phys.~Rev.~C { 87},  034609  (2013).

\bibitem{weppner} S.P: Weppner. R.B. Penney, G.W. Diffendale and G. Vittorini, Phys.~Rev.~C { 80}, 034608  (2009).

\bibitem{Koning-03} A.~J. Koning and J.~P. Delaroche, Nucl. Phys. { A713}, 231 (2003).

\bibitem{Cravo-16} E. Cravo, R. Crespo, and A. Deltuva, Phys.~Rev.~C { 93}, 054612 (2016).

\bibitem{bohr-motelson}
A. Bohr and B.~R. Motelson, {\em Nuclear Structure} (World Scientific,
  Singapore, 1998).

\bibitem{tamura:cex}
T. Tamura, Rev. Mod. Phys. { 37},  679  (1965).

\bibitem{thompson:88}
I.~J. Thompson, Computer Physics Reports { 7},  167  (1988).

\bibitem{Pham:76} Dinh-Liem Pham, Le Journal de Physique Lett. { 37}, L-67 (1976). 
\bibitem{Karban:69} O. Karban, J. Lowe, P.D. Graves, V. Hnizdo, Nucl. Phys { A133}  255 (1969).
\bibitem{Hannen} V.M. Hannen et al, Phys.~Rev.~C { 67}, 054320 (2003). 
\bibitem{Bob-overlaps} R.~B.~Wiringa, {\tt https://www.phy.anl.gov/theory/research/overlaps/}

\bibitem{wiringa:95a}
R.~B. Wiringa, V.~G.~J. Stoks, and R. Schiavilla, Phys.~Rev.~C { 51},  38
  (1995).

\bibitem{machleidt:01a}
R. Machleidt, Phys.~Rev.~C { 63},  024001  (2001).

\bibitem{amos:03a}
K. Amos, L. Canton, G. Pisent, J. Svenne, D. van der Knijff, Nucl. Phys. A
  728  (2003) 65.


\end{thebibliography}


\end{document}